\documentclass[12pt]{article}
\usepackage{amsthm,amssymb,amsmath}
\usepackage[british]{babel}

 1 
1

\newcommand{\C}{\ensuremath{\mathbb{C}}}

\newcommand{\bt}{{\boldsymbol{t}}}
\newcommand{\ba}{{\boldsymbol{a}}}

\newcommand{\bu}{{\boldsymbol{u}}}
\renewcommand{\d}{\operatorname{d}}
\newtheorem{teh}{Theorem}

\newtheorem{lem}{Lemma}

\newcommand{\be}{\begin{equation}}
\newcommand{\ee}{\end{equation}}
\begin{document}

\title{\sc Quasiconformal Mappings and Solutions of the
 Dispersionless KP hierarchy
\thanks{ B. Konopelchenko  is supported in part by COFIN
2000 "Sintesi".  L. Martinez Alonso is supported by the \emph
{Fundaci\'{o}n Banco Bilbao Vizcaya Argentaria} . E. Medina is
supported in part by CICYT proyecto PB98--0821 }}
\author{B. Konopelchenko$^{1}$ , L. Mart\'{\i}nez Alonso$^{2}$ and E.
Medina$^{3}$
 \\
\emph{ $^1$Dipartimento di Fisica, Universita
di Lecce and Sezione INFN}\\ \emph{73100 Lecce, Italy} \\
\emph{$^2$Departamento de F\'{\i}sica Te\'{o}rica II, Universidad
Complutense }\\\emph{ E-28040 Madrid, Spain}
 \\
\emph{ $^3$Departamento de Matem\'{a}ticas, Universidad de C\'{a}diz}\\
\emph{E-11510, Puerto Real, C\'{a}diz, Spain}\\
}
\date{} \maketitle
\begin{abstract}

A $\bar{\partial}$-formalism for studying dispersionless
integrable hierarchies is applied to the dKP hierarchy.
Connections with the theory of quasiconformal mappings on the
plane are described and some clases of explicit solutions of the
dKP hierarchy are presented.

\end{abstract}

\vspace*{.5cm}

\begin{center}\begin{minipage}{12cm}

\end{minipage}
\end{center}
\newpage

\section{Introduction}

Dispersionless or quasiclassical integrable hierarchies form an
important part of the theory of integrable systems . They are the
main ingredients of various approaches to different problems
arising in physics and applied mathematics (see e.g.
\cite{1}-\cite{10}) .  In this sense, it is worth-mentioning their
connections \cite{11,12} with some classical problems of the
theory of conformal maps. Furthermore,  we have recently proposed
a $\bar{\partial}$-method \cite{13}-\cite{14} for studying
dispersionless integrable hierarchies which reveals an intimate
relation between these hierarchies and the theory of
quasiconformal mappings \cite{15}-\cite{18}.

 The basic element of
our analysis is the nonlinear $\bar{\partial}$-equation
\begin{equation}\label{01}
S_{\bar{z}}=W\Big(z,\bar{z},S_{z}\Big),
\end{equation}
where $z\in\C$, $S(z,\bar{z},\bt)$ is a complex-valued function
depending on an infinite set $\bt$ of parameters (times),
$S_{\bar{z}}:=\frac{\partial S}{\partial
\bar{z}},\;S_{z}:=\frac{\partial S}{\partial z}$ and $W$ is an
appropriate function of $z,\bar{z}$ and $S_z$.

 As a consequence
of \eqref{01} the first-order derivatives of $S$ with respect to
the $\bt$ parameters satisfy the family of Beltrami equations
\begin{equation}\label{02}
f_{\bar{z}}=\mu(z,\bt)f_{z},
\end{equation}
where
\begin{equation}\label{03}
\mu :=W'(z,\bar{z}, S_z),\quad W'= W_{\xi}(z,\bar{z},\xi).
\end{equation}
This fact provides us with the link between the
$\bar{\partial}$-method and the theory of quasiconformal
mappings, in which Beltrami equation is of fundamental
importance. It should be mentioned that equation \eqref{01}, in
turn, is also well-known in the theory of quasiconformal mappings
(see e.g. \cite{19}-\cite{20}).

Our method is based on the determination of solutions of
\eqref{01} by means of the classical schemes for solving
first-order PDE's of Hamilton-Jacobi type . The objective of the
present paper is to illustrate our approach by presenting some
exact explicit solutions of the dKP hierarchy. We also prove that
the simplest of this solutions (Example 1 of Section 4) can not
be reached by the standard methods based on the hodograph
transformation technique \cite{3}-\cite{4}.

\section{The dKP hierarchy}

The dKP hierarchy can be introduced as the following
\emph{classical version} of the Lax-pairs equations of the
standard KP theory \cite{1}-\cite{9}
\begin{equation}\label{1}
\frac{\partial z}{\partial t_n}=\{\Omega_n,z\},\quad
 \Omega_n(p,\bt):=(z^n)_+,\quad n \geq 1.
\end{equation}
Here $z=z(p,\bt)$ is a complex function depending on a complex
variable $p$ and an infinite  set $\bt:=(t_1,t_2,\ldots)$ of
complex parameters, which is assumed to posses a Laurent
expansion of the form
\begin{equation}\label{2}
z=p+\sum_{n\geq 1}\frac{a_n(\bt)}{p^n},
\end{equation}
near $p\rightarrow\infty$. We denote by $(z^n)_+$ the polynomial
part of the expansion of $z^n$ in powers of $p$
\[
(z)_+=p,\quad (z^2)_+=p^2+2a_1,\quad (z^3)_+=p^3+3p\,a_1+3a_2,
\]
and the Poisson bracket is defined as
\[
  \{F,G\}:=\frac{\partial F}{\partial p}\frac{\partial G}{\partial
  x}-\frac{\partial F}{\partial x}\frac{\partial G}{\partial
  p},\quad x:=t_1.
\]
The compatibility conditions for \eqref{1} are of the form
\begin{equation}\label{3}
\frac{\partial \Omega_m}{\partial t_n}-\frac{\partial
\Omega_n}{\partial t_m}+\{\Omega_m, \Omega_n \}=0,\quad m\neq n,
\end{equation}

Two interesting examples of nonlinear equations of the dKP
hierarchy are:
\begin{description}
\item{1)}
 For $n=2$, equation \eqref{1} leads to the Benney
moment equations
\begin{equation}\label{4}
\frac{\partial a_{n+1}}{\partial t}+\frac{\partial
a_{n+2}}{\partial x}+na_{n}\frac{\partial a_1}{\partial x}=0,\quad
t:=-2t_2.
\end{equation}
\item{2)}
The compatibility equations \eqref{3} for $n=2$ and $m=3$  imply
the dKP equation (Zabolotskaya--Khokhlov equation)
\begin{equation}\label{5}
(u_t-\frac{3}{2}uu_x)_x=\frac{3}{4}u_{yy},\quad u:=2a_1,\;
t:=t_3,\; y:=t_2.
\end{equation}
\end{description}

From \eqref{3} it follows \cite{6}  that for any solution
$z=z(p,\bt)$ of the dKP hierarchy, there exists an associated
function $S=S(z,\bt)$, such that
\begin{equation}\label{6}
\frac{\partial S(z,\bt)}{\partial
t_n}=\Omega_n(p(z,\bt),\bt),\quad n\geq 1.
\end{equation}
Here $p=p(z,\bt)$ is obtained  by inverting the solution
$z=z(p,\bt)$. Without loss of generality it can be assumed that
$S$ has a Laurent expansion
\begin{equation}\label{7}
S(z,\bt)=\sum_{n\geq 1} z^n t_n+\sum_{n\geq
1}\frac{S_n(\bt)}{z^n},\quad z\in\Gamma,
\end{equation}
on a certain circle  $\Gamma=\{z:|z|=r\}$. Observe that by
setting $n=1$ in \eqref{6} and using \eqref{7} one characterizes
$p=p(z,\bt)$ in the form
\begin{equation}\label{8}
 p=\frac{\partial S(z,\bt)}{\partial
x}=
 z+\sum_{n\geq 1}\frac{b_n(\bt)}{z^{n}}, \quad b_n:=\frac{\partial S_n}{\partial
 x},
\end{equation}
Reciprocally, given a function $S=S(z,\bt)$ which satisfies
\eqref{6} and \eqref{7},  it can be proved \cite{6} that the
inverse function $z=z(p,\bt)$ of the function $p=p(z,\bt)$ of
\eqref{8}  determines a solution of the dKP hierarchy.

Henceforth, functions verifying conditions \eqref{6} and \eqref{7}
will be referred to as {\em $S$-functions} of the dKP hierarchy.
They turn out to be  related to the $\tau$-functions \cite{6}
according to
\[
S(z,\bt)=\sum_{n\geq 1} z^n t_n+\sum_{n\geq
1}\frac{1}{n\,z^n}\frac{\partial \ln\tau(\bt)}{\partial t_n}
,\quad z\in\Gamma.
\]
 We notice that  the system \eqref{6} for a dKP  $S$-function
constitutes a set of compatible Hamilton-Jacobi type equations
\begin{equation}\label{hj}
\frac{\partial S}{\partial t_n}=\Omega_n\Big(\frac{\partial
S}{\partial x},\;\bt\Big),\quad n\geq 2,
\end{equation}
which represents the \emph{semiclassical limit} of the linear
system for the wave function of the standard KP hierarchy.

Several methods for constructing solutions of the dKP hierarchy
through $S$-functions have been devised \cite{3,4,6,8}. We will
be here concerned with the $\bar{\partial}$-method proposed in
\cite{13}-\cite{14}, which aims to characterize $S$-functions by
means of solutions of $\bar{\partial}$-equations of the form
\eqref{01}. A main feature of this approach is that the
symmetries of \eqref{01} (first order variations $f:=\delta S$)
are determined by the family of Beltrami equations
\eqref{02}-\eqref{03}. This property implies, in particular, that
all the first-order derivatives $\frac{\partial S}{\partial t_n}$
of a solution of \eqref{01} satisfy \eqref{02}.

Given any fixed local solution $f$ of a Beltrami equation, if $f$
has non-zero Jacobian at a certain point $z_0$, then all the
smooth local solutions $F$ near $z_0$ are given by analytic
functions $F=F(f)$ of $f$ .  This property suggests that, under
\emph{appropriate conditions}, solutions $S$ of a
$\bar{\partial}$-equation \eqref{01} which admit expansions of
the form \eqref{7}, satisfy conditions \eqref{hj} as well.
Therefore, they provide $S$-functions for the dKP hierarchy. In
what follows a rigorous basis for this scheme is proposed.

\section{Quasiconformal mappings}

Quasiconformal mappings are a natural and very rich extension of
the concept of conformal mappings.  For the sake of convenience we
remind here some of their basic properties (see e.g.
\cite{13}-\cite{18}).

Let $\mu=\mu(z)$ be a measurable function  on a domain $G$ of the
complex plane such that for some $0<k<1$ it verifies $|\mu(z)|<k$
almost everywhere in $G$ . Then, a function $f=f(z)$ is said to be
a \emph{quasiconformal mapping (qc-mapping) with complex
dilatation $\mu$} in $G$ if
\begin{description}
\item[i)] $f$ is a homeomorphism $f:\,D\rightarrow D'$
\item[ii)] $f$ is a generalized solution of the linear Beltrami equation
\begin{equation}\label{12}
f_{\bar{z}}=\mu f_{z},
\end{equation}
on $D$, with locally square-integrable partial derivatives
$f_{\bar{z}}$ and $f_{z}$.
\end{description}

The properties of solutions of the Beltrami equation \eqref{12}
are rather well studied (see e.g. \cite{16}). Some of them are
particularly important for our discussion. Before presenting these
results we introduce the Calder'on-Zygmund operator \cite{16}
\begin{equation}\label{13}
(Th)(z):=\frac{1}{2\pi\imath}\int\!\!\int_{\C}\frac{h(z')}{(z'-z)^2}\d
z' \wedge \d \bar{z}',
\end{equation}
where the integral is taken in the sense of the Cauchy principal
value. Then one has the following fundamental result  \cite{16}:

\begin{teh}
For any $p\geq 2$ the operator $T$ defines a bounded operator in
$L^p(\C)$. Moreover, $||T||_p$ is continuos with respect to $p$
and satisfies
\[
\lim_{p\rightarrow 2}||T||_p=1.
\]
\end{teh}

As a consequence of this theorem it follows that for any $0\leq
k<1$ there exists $\delta(k)>0$ such that
\[
k||T||_p<1,
\]
for all $2<p<2+\delta(k)$.

The next theorem provides us with  a  property of qc-mappings
which will be useful in the $\bar{\partial}$-method .

\begin{teh} Let $\mu$ be a measurable function with compact support
inside the circle $|z|<R$ and such that $||\mu||_{\infty}<k<1$.
Then, for any  $p>2$ such that $k||T||_p<1$, it follows that the
only generalized solution of  Beltrami equation \eqref{2}
verifying
\begin{equation}\label{14}
f(z)=O(\frac{1}{z}),\quad z\rightarrow\infty,
\end{equation}
and $f_{{\bar z}},f_{z}\in L^p(\C)$ is $f\equiv 0$.
\end{teh}

This result is the uniqueness part of the existence theorem for
the so-called normal solutions of Beltrami equations
\cite{15}-\cite{17}. Its proof relies on the fact that the
operator $T$ represents the action of
$\partial_z\partial_{\bar{z}}^{-1}$ on $L^p(\C)$, so that under
the hypothesis of the theorem we have that Beltrami equation for
$f$ becomes the integral equation
\[
\phi-\mu T\phi=0,\quad \phi:=f_{\bar{z}},
\]
on $L^p(\C)$. Thus, by taking into account that $||\mu T||_p\leq
k||T||_p<1$, we have $\phi\equiv 0$ and then, in view of
\eqref{14}, $f\equiv 0$.

Let us go back to our $\bar{\partial}$-equation \eqref{01} and let
us assume that $W(z,\bar{z},S_z)$ vanishes for all $z$ outside a
certain circle $\Gamma=\{z:|z|=r\}$. Suppose that  we are able to
find a solution $S=S(z,\bar{z},\bt)$ of \eqref{01} in the disk
$D=\{z:|z|<r\}$ with a boundary value
$S|_{\Gamma}:=S(z,\frac{r^2}{z},\bt)$ of the form \eqref{7}, and
such that for a certain $0<k<1$ the set
\[
\Omega:=\{\bt:\sup_{z\in D}|W'(z,\bar{z},S_z(z,\bt)|\leq k\},
\]
is not empty. In this case we can apply Theorem 2 to the Beltrami
equation \eqref{02}-\eqref{03}. Moreover, by taking into account
that
\[
\frac{\partial S}{\partial t_n}=z^n+\sum_{m\geq 1}\frac{\partial
S_m(\bt)}{\partial t_n}z^{-m} ,\quad z\in\Gamma,
\]
these functions can be continuously extended from $D$ to analytic
functions outside $\Gamma$. Thus, they become solutions of
\eqref{02}-\eqref{03} on the whole complex plane. On the other
hand, it is clear that
\[
\frac{\partial S}{\partial t_n}-(z^n)_+= \frac{\partial
S}{\partial t_n}-\Omega_n\Big(\frac{\partial S}{\partial
x},\;\bt\Big)=O(\frac{1}{z}),\quad z\rightarrow\infty,
\]
so that from Theorem 2 we conclude that \eqref{6} is satisfied
and, consequently, $S=S(z,\bar{z},\bt)$ determines an $S$-function
of the dKP hierarchy for $\bt\in \Omega$.

\section{Solutions of the dKP hierarchy }

In order to construct explicit solutions of the dKP hierarchy we
consider  $\bar{\partial}$-equations of the form
\begin{equation}\label{15}
 S_{\bar{z}}=\theta(r-|z|)V\Big(z,\bar{z},S_z\Big),
\end{equation}
where $r>0$, $\theta(\xi)$ is the usual Heaviside function and $V$
is an analytic  function of $z$, $\bar{z}$ and $S_z$. Our scheme
of solution is as follows
\begin{description}
\item[1)] Firstly, we generate solutions $S=S(z,\ba)$ of \eqref{15}
\begin{equation}\label{16}
S_{\bar{z}}=V\Big(z,\bar{z},S_z\Big),\quad |z|<r,
\end{equation}
depending on a set of free parameters $\ba:=(a_0,a_1,\ldots)$.
\item[2)] Then we select those solutions whose boundary value
on $\Gamma=\{z:|z|=r\}$ is of the form \eqref{7}.
\end{description}

Equation \eqref{16} is a PDE of Hamilton-Jacobi type , so that
several powerful methods for generating solutions are available.
For example, if $V=V(S_z)$ depends only on $S_z$ then  \eqref{16}
implies
\[
 m_{\bar{z}}=V_m(m)m_z,\quad m:= S_z.
\]
This equation can be solved at once by applying the methods of
characteristics. So the general solution \eqref{16} is implicitly
characterized by
\begin{equation}\label{17}
\begin{gathered}
S=V(m)\bar{z}+m\,z-f(m),\\ \\V_m\bar{z}+z=f_m(m),
\end{gathered}
\end{equation}
where $f=f(m)$ is an arbitrary function. Notice that according to
the second equation in \eqref{17}, we have
\[
f_m(m_0)=z,\quad m_0:=m(z,\bar{z})|_{\bar{z}=0},
\]
so that $f_m(m_0)$ is the inverse function of $m_0=m_0(z)$.


As we are aiming to get explicit solutions some simplifying
assumptions are required. For instance, we consider cases in which
only a finite set of $N+1$ parameters $\ba=(a_0,a_1,\ldots,a_N)$
are involved. Therefore we are facing the problem of selecting
solutions $S$ in which no terms $z^n t_n$ with $n>N$ in \eqref{7}
appear. Other type of solutions of the $\bar{\partial}$-equation
would have time-parameters $t_n\;(n>N)$ which are functions of
$(t_1,\ldots,t_N)$ and no solution of the dKP hierarchy would
arise in that way.

 In order to explore the
possible favorable cases, let us consider the class of
$\bar{\partial}$-equations \eqref{16} of the form
\begin{equation}\label{19}
S_{\bar{z}}=\bar{z}^s\sum_{m\geq 0}^{M} p_m(z)(S_z)^m,\quad |z|<r,
\end{equation}
where $s\geq 0$, $M\geq 2$, the coefficients $p_m=p_m(z)$ are
polynomials in $z$ and $p_M\not\equiv 0$. Let us look for a
series solution of \eqref{19} of the form
\begin{equation}\label{20}
S=\sum_{n\geq 0}c_n(z)\bar{z}^{n(s+1)},
\end{equation}
with $c_0$ being set as an arbitrary $N$-degree polynomial
($N\geq2$)
\begin{equation}\label{20b}
c_0(z)=\sum_{n=0}^{N} a_nz^n,\quad a_N\neq 0.
\end{equation}
Substitution of \eqref{20} in \eqref{19} provides the recursion
relation
\begin{equation}\label{20a}
c_{n+1}=\frac{1}{(n+1)(s+1)}\sum_{m\geq
0}^{M}p_m(z)\Big(\sum_{r_1+\cdots r_m=n}c'_{r_1}\cdots
c'_{r_m}\Big),\quad n\geq 0,
\end{equation}
which shows that all coefficients $c_n(z)$ of \eqref{20} are
polynomials. We need to know their degree in order to be able to
examine the form of $S$ on the boundary $|z|=r$.

\begin{lem}
If the degrees of the coefficients $p_m$ in \eqref{19} verify
\[
deg(p_m)\leq (M-m)(N-1),\quad m=0,1,\ldots,M,
\]
then
\begin{equation}\label{21}
\mbox{deg}(c_n)=n[M(N-1)-N]+N,\quad n\geq 0.
\end{equation}
\end{lem}

\noindent {\bf Proof}

We apply the induction principle. It is obvious that \eqref{21} is
true for $n=0$. Suppose now that it holds for $n'\leq n$, and
consider the terms in the expression \eqref{20a} for $c_{n+1}$.
By taking into account that $r_1+\cdots+r_m=n$ we get
\begin{align*}
deg(p_m\, c'_{r_1}\cdots c'_{r_m})&=
\sum_{i=1}^m\Big[r_i\Big(M(N-1)-N\Big)+N-1\Big]+deg(p_m)
\\&=n\Big[M(N-1)-N\Big]+m(N-1)+deg(p_m)\\&\leq
n\Big[M(N-1)-N \Big]+M(N-1)\\&=(n+1)\Big[M(N-1)-N]+N.
\end{align*}
Moreover, as $p_M$ is a non-zero constant, it is clear that the
corresponding terms in \eqref{20a} verify
\begin{align*}
deg(p_M\, c'_{r_1}\cdots
c'_{r_M})&=n\Big[M(N-1)-N\Big]+M(N-1)\\&=(n+1)\Big[M(N-1)-N\Big]+N.
\end{align*}
Thus, one  concludes that \eqref{21} is verified for $c_{n+1}$ too
and, therefore, the statement is proved (QED).

If we assume that the series \eqref{20} converges for a certain
$r\geq 0$, then under the hypothesis  of the above lemma the
continuous extension of $S$ to the boundary $|z|=r$ is of the form
\begin{equation}\label{22}
S=\sum_{n\geq 0}r^{2n(s+1)}\frac{c_n(z)}{z^{n(s+1)}},\quad
\frac{c_n(z)}{z^{n(s+1)}}=O\Big(z^{d_n}\Big),
\end{equation}
where
\begin{equation}\label{23}
d_n=n[M(N-1)-N-s-1]+N.
\end{equation}
Since the solution corresponding to \eqref{20b} depends on $N+1$
free parameters $(a_0,\ldots,a_N)$, only those cases for which
$d_n\leq N$ for all $n\geq 0$ are of interest. From \eqref{23} it
is obvious that this happens only if
\begin{equation}\label{in}
N\leq\frac{M+s+1}{M-1}.
\end{equation}

For example if we set $s=0$, this means that we have only the
three possibilities exhibited in the following table:

\vspace{.2cm}

\noindent \hfil\vbox{\hbox{\vbox{\offinterlineskip
\halign{&\vrule#&\strut\quad#\hfil\quad\cr
\noalign{\hrule}
height2pt&\omit&&\omit&\cr
 &$(M,N)$ && $V(z,S_z)$ && $S|_{\bar{z}=0}$ &\cr
   height2pt&\omit&&\omit&\cr
   \noalign{\hrule}
   height2pt&\omit&&\omit&\cr
 &(2,2) && $\alpha(S_z)^2+(\sum_{i=0}^1\beta_iz^i)S_z
   +\sum_{i=0}^2\gamma_i z^i$ && $\sum_{i=0}^2 a_iz^i$ &\cr
   height2pt&\omit&&\omit&\cr
   \noalign{\hrule}
   height2pt&\omit&&\omit&\cr
 &(3,2)&& $\alpha(S_z)^3+(\sum_{i=0}^1\beta_i z^i)(S_z)^2
   +(\sum_{i=0}^2\gamma_i z^i)S_z+\sum_{i=0}^3\eta_i z^i$ && $\sum_{i=0}^2 a_iz^i$  &\cr
   height2pt&\omit&&\omit&\cr
   \noalign{\hrule}
   height2pt&\omit&&\omit&\cr
 &(2,3)&& $\alpha(S_z)^2+(\sum_{i=0}^2\beta_iz^i)S_z
   +\sum_{i=0}^4\gamma_i z^i$ &&  $\sum_{i=0}^3 a_iz^i$  &\cr
   height2pt&\omit&&\omit&\cr
\noalign{\hrule}}}}}\hfill

\vspace{.2cm}

 \vspace{.2cm}

\noindent {\bf Example 1.} The simplest case in the class $(2,2)$
corresponds to
\begin{equation}
 S_{\bar{z}}=\theta(1-|z|)(S_z)^2,
\end{equation}
with $S|_{\bar{z}=0}$ being a quadratic polynomial. This yields to

\[
S=\left\{ \begin{array}{l}
\frac{1}{2}\frac{(z-b)^2}{a-2\bar{z}}-c,\quad |z|\leq 1\\ \\
\frac{1}{2}\frac{z(z-b)^2}{az-2}-c,\quad |z|\geq 1.
\end{array}
\right.
\]
Notice that the regularity of $S$ inside the unit circle requires
\begin{equation}\label{24}
|a|>2.
\end{equation}

On the boundary $|z|=1$ we have
\[
S=\frac{1}{2a}z^2+(\frac{1}{a^2}-\frac{b}{a})z+
\frac{2}{a^3}+\frac{b^2}{2a}-\frac{2b}{a^2}-c+O(\frac{1}{z}).
\]
So, in order to fit with the required form of an $S$-function of
the  dKP hierarchy we have to identify
\[
x=\frac{1}{a^2}-\frac{b}{a},\quad t_2=\frac{1}{2a},\quad
c=\frac{2}{a^3}+\frac{b^2}{2a}-\frac{2b}{a^2}.
\]

On the other hand, the complex dilatation for the corresponding
Beltrami equation \eqref{01}-\eqref{02} is given by
\begin{equation}\label{25}
\mu(z,\bar{z}):=2\theta(1-|z|)\frac{z-b}{a-2\bar{z}}.
\end{equation}
We observe that the following bound follows
\[
|\mu(z,\bar{z})|<2\frac{|b|+1}{|a|-2},\quad z\in \C.
\]
In this way, for any $0<k<1$ we have $|\mu(z)|\leq k$ provided
$k|a|>2(|b|+k+1)$. Thus, there is a non empty domain in the space
of parameters on which the Beltrami equation \eqref{01}-\eqref{02}
satisfies the conditions assumed in our discussion of Section 3.

It follows that
\[
p:=S_x=\frac{z^2-4t_2z+2(x+4t_2^2)}{z-4t_2},\quad |z|=1,
\]
so one gets the following solution of the $t_2$-flow (Benney flow)
of the dKP hierarchy
\begin{equation}\label{ss}
z=\frac{p}{2}+2t_2+\sqrt{(\frac{p}{2}-2t_2)^2-2x-8t_2^2}.
\end{equation}

We notice that this solution depends on the time-parameters
trough the pair of functions $u_1:=2t_2,\; u_2:=-2x-8t_2^2$.
Indeed, we can rewrite it as
\[
z=\frac{p}{2}+u_1+\sqrt{(\frac{p}{2}-u_1)^2+u_2}.
\]
In case this solution could be obtained by the hodograph methods
\cite{3}-\cite{4}, it would correspond to a reduction of the dKP
hierarchy in which the functions $\bu=(u_1,u_2)$ would satisfy a
diagonalizable hydrodynamic type system with Riemann invariants
provided by the zeros of the function $\frac{\partial
z(p,\bu)}{\partial p}$. But this function has no zeros for
$u_2\neq 0$ as
\[
\frac{\partial z(p,\bu)}{\partial
p}=\frac{\frac{p}{2}-u_1+\sqrt{(\frac{p}{2}-u_1)^2+u_2}}{2\sqrt{(\frac{p}{2}-u_1)^2+u_2}}.
\]
Therefore, we conclude that the solution \eqref{ss} of the dKP
hierarchy can not be reached from the hodograph technique
approach.

\vspace{.2cm}

\noindent {\bf Example 2.} The $\bar{\partial}$-equation
\begin{equation}
 S_{\bar{z}}=\theta(1-|z|)(S_z)^3,
\end{equation}
with $S|_{\bar{z}=0}$ being a quadratic polynomial, is an example
of the $(3,2)$ case. One finds the $S$-function
\[
S=\bar{z}m^3+(z-b)m-\frac{a}{2}m^2-c,\quad |z|<1,
\]
where
\[
m=\frac{1}{6\bar{z}}\Big(a-\sqrt{a^2-12(z-b)\bar{z}}\Big).
\]
The defining relations for the dKP parameters $(x,t_2)$ are
\begin{equation}\label{26}
x=\frac{b}{6}\Big(\sqrt{a^2-12}-a\Big),\;
t_2=\frac{1}{108}\Big(18a-a^3+(a^2-12)^{3/2}\Big).
\end{equation}
The corresponding solution of the $t_2$-flow of the dKP hierarchy
can be written as
\begin{align*}
z&=a(a-\sqrt{a^2-12})\frac{p}{12}+\frac{3x}{\sqrt{a^2-12}}+\frac{1}{12}
\Big[\Big( a(a-\sqrt{a^2-12})p+\frac{36x}{\sqrt{a^2-12}}
\Big)^2\\\\
&-12\Big((a-\sqrt{a^2-12})p+\frac{3(a+\sqrt{a^2-12})x}{\sqrt{a^2-12}}
\Big)^2\Big]^{\frac{1}{2}},
\end{align*}
where $a=a(t_2)$ is to be obtained from \eqref{26}.

\vspace{.2cm}

\noindent {\bf Example 3.} The class $(2,3)$ is the most
interesting one as it provides solutions of the dKP hierarchy
depending of $(x,t_2,t_3)$. Let us consider
\begin{equation}
 S_{\bar{z}}=\theta(1-|z|)(S_z)^2,
\end{equation}
with a cubic polynomial $S|_{\bar{z}=0}$. If we take
$m_0:=S_z|_{\bar{z}}=az^2+bz+c$, then from \eqref{17} we get
\begin{equation}\label{27}
\begin{gathered}
f(m)=-\frac{b}{2a}m+\frac{1}{12
a^2}(4am+b^2-4ac)^{\frac{3}{2}}+d\\
=-\frac{b}{2a}m+\frac{1}{12 a^2}(4a\bar{z}m+2az+b)^3+d,
\end{gathered}
\end{equation}
and
\begin{equation}\label{28}
m=\frac{1}{8\bar{z}^2}\Big(\frac{1}{a}-4(z+\frac{b}{2a})\bar{z}-\sqrt{\frac{4}{a}
(\frac{b^2}{a}-4c)\bar{z}^2-\frac{8}{a}
(z+\frac{b}{2a})\bar{z}+\frac{1}{a^2}}\;\;\Big).
\end{equation}
Hence we have
\begin{equation}\label{29}
S=(z+\frac{b}{2a})m+\bar{z}m^2-\frac{1}{12
a^2}(4a\bar{z}m+2az+b)^3-d,\quad |z|<1.
\end{equation}
It is clear that in order to ensure $S$ to be continuous we have
to require \cite{14}
\[
\frac{4}{a} (\frac{b^2}{a}-4c)\bar{z}^2-\frac{8}{a}
(z+\frac{b}{2a})\bar{z}+\frac{1}{a^2}\neq 0,\quad |z|<1.
\]

Notice also that $S$ is regular at the origin as
\[
\lim_{z\rightarrow 0}m=c.
\]

Let us outline the calculation of $u=-2\frac{\partial
S_1}{\partial x}$. We need to compute the first few terms of the
expansion of $S$ on $|z|=1$. To do that it is helpful to use the
following identity
\begin{equation}\label{30}
S_z= m-\Big(\frac{m}{z}\Big)^2,\quad |z|=1.
\end{equation}
Thus, from the expansion of $m$ on $|z|=1$ and by setting
\[
S_z=3t_3z^2+2t_2z+x-\frac{S_1}{z^2}+\ldots,\quad |z|=1,
\]
in \eqref{30}, we get
\begin{equation*}
\begin{gathered}
3t_3=-\frac{3}{4}+\frac{3}{8a}+\frac{1}{32
a^2}\Big((1-8a)^{3/2}-1\Big),\quad\quad
2t_2=\frac{b}{8a^2}(1-4a-\sqrt{1-8a}),\\\\
x=-\frac{b^2}{8a^2}\Big(1+\frac{4a-1}{2\sqrt{1-8a}}\Big)+
\frac{c}{4a}(1-\sqrt{1-8a}),
\end{gathered}
\end{equation*}
and
\begin{equation}\label{32}
S_1=\frac{\Big(2b^2+(1-8a)c\Big)^2}{(1-8a)^{5/2}}.
\end{equation}
From these expressions it can be computed that $u=-2\frac{\partial
S_1}{\partial x}$ is given by ($t:=t_3,y:=t_2$)
\begin{equation}\label{33}
u=\frac{4\Big(5-12t+4\sqrt{1-12t}\Big)^2\Big((-1+12t)x-4y^2\Big)}
{3(1+4t)(1-12t+2\sqrt{1-12t})^3},
\end{equation}
which satisfies \eqref{5}. We notice that according to \cite{10},
\eqref{33} belongs to the class of solutions of the dKP equation
which yield  Einstein-Weyl structures  conformal to Einstein
metrics .

More general cases corresponding to \eqref{19} with $s\neq 0$ and
$z$-dependence on its right-hand side will be considered
elsewhere \cite{21}.

\noindent
{\bf Acknowledgements}

B. Konopelchenko is grateful to the organizers  of the programme
"Integrable Systems" for the support provided.   L. Martinez
Alonso wish to thank the \emph{Fundaci\'{o}n Banco Bilbao Vizcaya
Argentaria} for supporting his stay at Cambridge University as a
BBV visiting professor.

\end{document}